\documentclass[preprint,aps,nofootinbib]{revtex4}
\usepackage{latexsym,amsmath,bm,epsfig}

\newcommand{\x}{{\vec x}}

\newcommand{\p}{{\vec p}}

\renewcommand{\d}{\partial}
\newcommand{\<}{\langle}
\renewcommand{\>}{\rangle}
\newcommand{\D}{{\mathcal{D}}}
\newcommand{\stru}{\rule[-.25in]{0in}{.25in}}
\newcommand{\struu}{\rule[-.325in]{0in}{.325in}}
\newcommand{\half}{{\textstyle\frac12}}
\begin{document}

\preprint{CU-TP-1081}
\preprint{INT-PUB 02-56}
\vspace*{1cm}
\title{On the equivalence between the Boltzmann equation\\ and classical
field theory at large occupation numbers}

\author{A.~H.~Mueller}
\affiliation{Departmet of Physics \\ 
Columbia University\\ 
New York, NY 10027}

\author{D.~T.~Son} 
\affiliation{Institute for Nuclear Theory\\ 
University of Washington\\
Seattle, WA 98195-1550\\}

\begin{abstract}
We consider a system made up of exictations of a neutral scalar field,
$\phi$, having a $\lambda\phi^4$ interaction term.  Starting from an
ensemble where the occupation number $f$ is large, but $\lambda f$ is
small, we develop a classical field theory description of the
evolution of the system toward equilibrium.  A Boltzmann equation
naturally emerges in this description and we show by explicit
calculation that the collision term is the same as that coming from
elastic scattering.  This shows the equivalence of a Boltzmann
equation description and a classical field theory description of the
same system.
\end{abstract}

\maketitle

\section{Introduction}
\label{sec1}
It is believed that in the very early stages after a heavy ion
collision, gluon occupation numbers are as large as $1/\alpha$,
$f_g\sim1/\alpha$~\cite{nuc1,nuc2,nuc3,nuc4,nuc5}.  As the QCD
quark-gluon plasma evolves toward equilibrium, occupation numbers
decrease until $f_g\sim1$ at equlibrium.  Understanding the evolution
of dense quark-gluon matter toward equilibrium is both practically
important and theoretically challenging.  So long as $f_g\ll1/\alpha$
the Boltzmann equation, with elastic and inelastic collision terms and
quantum-statistical factors, should furnish a systematic theoretical
framework for this problem~\cite{BMSS}, although this equation cannot
be expected to be useful at the earliest times after the collision
when $f_g\sim1/\alpha$.  On the other hand, when occupation numbers
are large one can expect classical field theory to apply, and there is
currently an interesting program studying the early stages after a
heavy ion collision using classical Yang-Mills equations with an
initial condition fixed by the McLerran-Venugopalan saturation
model~\cite{KV1,KV2}.  Thus one might expect that classical field
theory should be a good theoretical framework for studying a dense
gluon system up to times just before equilibration occurs.  It would
seem then that classical field theory and the Boltzmann equation are
equivalent frameworks for studying dense non-equilibrium QCD in the
period where $1\ll f_g\ll1/\alpha$ holds for the important phase space
region of the system.

While this equivalence have been studied previously, most notably in
connection with wave turbulence~\cite{ZLF}, it is less familiar in the
context of the early stages of heavy ion collision.  We thus explore
this equivalence in detail in this paper.  To simplify the discussion
we deal with a $\phi^4$ field theory rather than with QCD, but we
believe our argumentation should work equally well for Yang-Mills
theories.  Our goal is to derive the collision term in the Boltzmann
equation using only classical field theory, and we do this explicitly
at lowest order in the coupling for the elastic-scattering part of the
collision term.  This part of the collision term, illustrated in
Fig.~(\ref{fig5}) below, has a contribution cubic in the occupation
number and a contribution which is quadratic.  Perhaps surprisingly we
are able to reproduce both of these parts from the classical theory,
and the reasons for this are discusssed briefly at the end of
Sec.~\ref{sec4}.  For inelastic parts of the collision term the
classical field description cannot be expected to give the complete
answer, even at the lowest nontrivial term in the coupling, but it
should give these portions having the maximum and maximum minus one
powers of the occupation number.

In Sec.~\ref{sec2} we review the description of a non-equilibrium
system, described by a $\phi^4$ field theory including the usual
doubling of the number of fields.  When the fields are large we show
how a classical field theory naturally replaces the quantum field
theory description of the system, and we introduce combinations of the
field variables to make the classical description transparent.

In Sec.~\ref{sec3} we introduce Green's functions in the classical
field theory and write equations of motions for the Green's functions.
If the medium does not vary too rapidly we show how a Boltzmann-like
equation emerges with a ``collision term'' given by Green's functions
of the classical theory, depending of course on the initial ensemble
defining the system.

In Sec.~\ref{sec4} we identify the ``collision'' term derived in the
classical theory with the usual collision term given by elastic
scattering.  The identification is done by explicit calculation at
lowest nontrivial order in perturbation theory.

Our whole discussion is carried out under conditions where the
occupation number $f$ is large ($f\gg1$) while $\lambda f$ is small
($\lambda f\ll1$) where $\lambda$ is the usual $\phi^4$ coupling
constant.  We believe that higher order corrections in $\lambda f$ can
be done in the classical theory, but higher order in $\lambda$ will,
in general, be quantum.

\section{Describing a dense non-equilibrium system}
\label{sec2}

The system which concerns us here is made up of excitations of a
neutral scalar field of mass $m$ and with interactions described by a
$-\frac\lambda{4!}\phi^4$ interaction term in the Lagrangian density.
Suppose at time $t$ we wish to determine the expectation of some
observable, $O$, made of of field $\phi$.  For example, $O$ might be
$\phi^2(\x,t)$, $\phi(\x_1,t)\phi(\x_2,t)$,
$\phi^2(\x_1,t)\phi^2(\x_2,t)$ etc.  We may write
\begin{equation}\label{eq1}
  \< O\> = \int\!\D[\phi]\, O[\phi]\int\!\D[\phi_0]\,\rho[\phi_0]\,
  U^*[\phi,\phi_0,t-t_0]\,U[\phi,\phi_0,t-t_0]
\end{equation}
with
\begin{equation}\label{eq2}
  U[\phi,\phi_0,t-t_0]=\int\!\D[\phi(\tau)]\,e^{
  i\int_{t_0}^t\! L[\phi(\tau)]\, d\tau}
\end{equation}
where the functional integral on the right hand side of
Eq.~(\ref{eq2}) goes between fields $\phi_0(\x)$ at $t_0$ and
$\phi(\x)$ at $t$.  The functional $\rho[\phi_0]$ gives the initial
ensemble defining the system.  We make no assumption of an equilibrium
or near-equilibrium ensemble.  Later on we shall state some general
assumptions on $\rho$.

The separate functional integrals between $t_0$ and $t$ in $U$ and
$U^*$ is characteristic of any real-time formalism describing the time
evolution of a statistical system~\cite{LeBellac}.  Rather than
viewing $U^*U$ as determined by two separate functional integrals over
the single field $\phi$ one can introduce two fields, $\phi_-$ and
$\phi_+$, with Lagrangian
\begin{equation}\label{eq3}
  \mathcal{L}= \left[\frac12 (\d_\mu\phi_-)^2 - \frac12m^2\phi_-^2
  -\frac\lambda{4!}\phi_-^4\right] - \left[\frac12(\d_\mu\phi_+)^2
  -\frac12m^2\phi_+^2-\frac\lambda{4!}\phi_+^4\right]
\end{equation}
Now
\begin{equation}\label{eq4}
  U^*U = \int\!\D[\phi_-]\,\D[\phi_+]\, e^{i\int_{t_0}^t\!
  L[\phi_-,\phi_+]\,d\tau}
\end{equation}
with both $\phi_-$ and $\phi_+$ taking on values $\phi(\x)$ at $t$ and
$\phi_0(\x)$ at $t_0$.

It is convenient to rewrite $\mathcal{L}$ in Eq.~(\ref{eq3}) in terms
of new variables $\phi$ and $\pi$ defined by~\cite{ra}
\begin{eqnarray}
  &&\phi=\frac12(\phi_-+\phi_+)\,,\qquad \pi = \phi_--\phi_+\,,
    \label{eq5}\\
  &&\phi_-=\phi+\frac\pi2\,,\qquad\qquad \phi_+=\phi-\frac\pi2\,.
    \label{eq6}
\end{eqnarray}
One easily finds
\begin{equation}\label{eq7}
  \mathcal{L} = \d_\mu\phi\d_\mu\pi - m^2\phi\pi - \frac\lambda{3!}
  \left(\phi^3\pi+{\textstyle \frac14}\pi^3\phi\right)
\end{equation}
where one should be careful not to confuse the $\phi$ in
Eqs.~(\ref{eq5})--(\ref{eq7}) with the original field appearing in
Eqs.~(\ref{eq1})--(\ref{eq2}).  In what follows $\phi$ will always
denote the variable defined in Eq.~(\ref{eq5}).  Noting that $\pi=0$
at times $t_0$ and $t$, one can rewrite Eq.~(\ref{eq1}) as
\begin{equation}\label{eq8}
  \<O\> = \int\!\D[\phi]\,O[\phi]\int\!\D[\phi_0]\,\rho[\phi_0]
  \int\!\D[\phi(\tau)]\,\D[\pi(\tau)]\,e^{i\int_{t_0}^t\!
  L[\phi,\pi]\,d\tau}\,,
\end{equation}
with
\begin{equation}\label{eq9}
  L[\phi,\pi]=\int\!d^3x\,\mathcal{L}[\phi,\pi]\,.
\end{equation}

Our focus is on systems where $\phi_-$ and $\phi_+$ are large.  As we
shall see, $\phi$ is naturally large while $\pi$ is small, thus one
can neglect the $\pi^3\phi$ term in Eq.~(\ref{eq7}), which gives
\begin{equation}\label{eq10}
  \mathcal{L}_c = \d_\mu\phi\d_\mu\pi -m^2\phi\pi-\frac\lambda{3!}\pi\phi^3
\end{equation}
But now $\pi$ is simply a variable of constraint so that the
functional integral over $\pi$ in Eq.~(\ref{eq8}), with $L$ replaced
by $L_c$, gives
\begin{equation}\label{eq11}
  \prod_{\x,\tau} 2\pi\,\delta\left[(\Box+m^2)\phi(\x,\tau)+
  \frac\lambda{3!}\phi^3(\x,\tau)\right]\,,
\end{equation}
which means that only fields satisfying the classical equations of motion
\begin{equation}\label{eq12}
  (\Box+m^2)\phi = -\frac\lambda{3!}\phi^3
\end{equation}
contribute to the functional integral in Eq.~(\ref{eq8}).  In other
words, if one defines the evolution kernel
\begin{equation}\label{eq13}
  K[\phi,\phi_0,t-t_0] = \int\!\D[\phi(\tau)]\,\D[\pi(\tau)]\,
  e^{i\int_{t_0}^t\,L_c[\phi,\pi]\,d\tau}\,,
\end{equation}
where, as usual, $\phi_0$ and $\phi$ are the fields between which the
functional integral is taken, then $K$ is nonzero only for $\phi$
statisfying the field equation with the initial condition
$\phi(t_0,\x)=\phi_0(\x)$, i.e.,
\begin{equation}\label{eq14}
  \left[(\Box+m^2)\phi(\x,t)+\frac\lambda{3!}\phi^3(\x,t)\right]
  K[\phi,\phi_0,t-t_0] = 0\,,
\end{equation}
and
\begin{equation}\label{eq15}
  \lim_{t\to t_0}
  K[\phi,\phi_0,t-t_0] =\prod_{\x}\delta[\phi(\x)-\phi_0(\x)]\,,
\end{equation}
The classical nature of $K$ is manifested in Eqs.~(\ref{eq14}) and
(\ref{eq15}).  In terms of $K$ one can write $\<O\>$ as
\begin{equation}\label{eq16}
  \<O\>=\int\!\D[\phi]\,\D[\phi_0]\,O[\phi]\,K[\phi,\phi_0,t-t_0]\,
  \rho[\phi_0]\,.
\end{equation}
In fact, the functional integral~(\ref{eq13}) with the Lagrangian of
the type~(\ref{eq10}) provides a convenient starting point to a
diagrammatic approach for classical statistical systems~\cite{MSR}.

\section{Equations for the Green's functions}
\label{sec3}

In this section we derive equations for the Green's functions of a
system with Lagrangian given by Eq.~(\ref{eq10}).  We
begin our discussion with the Green's functions (propagators) of the
free theory corresponding to
\begin{equation}\label{eq17}
  \mathcal{L}_0 = \d_\mu\phi\d_\mu\pi-m^2\phi\pi\,,
\end{equation}
then we go on to the equations for the full (classical) theory
governed by the Lagrangian~(\ref{eq10}).  We limit ourselves in this
section to vacuum propagators, leaving the generalization to the
medium case to Sec.~\ref{sec4}.

\subsection{The free propagators}

Perhaps the easiest way to get the propagators for the
Lagrangian~(\ref{eq17}) is to go back to the $\phi_-$ and $\phi_+$
variables, in which case
\begin{equation}\label{eq18}
  \mathcal{L}_0 = \frac12(\d_\mu\phi_-)^2-\frac12m^2\phi_-^2-
  \frac12(\d_\mu\phi_+)^2+\frac12m^2\phi_+^2
\end{equation}
where $\phi_-$, $\phi_+$, $\phi$, and $\pi$ are related according to
Eqs.~(\ref{eq5}) and (\ref{eq6}).  Then
\begin{equation}\label{eq19}
  G^{(0)}_{--}(x) = \<0|T\phi_-(x)\phi_-(0)|0\> =
  \int\!\frac{d^4k}{(2\pi)^4}\,\frac i{k^2-m^2+i\epsilon}e^{-ik\cdot x}
\end{equation}
is the usual Feynman propagator.  In momentum space
\begin{equation}\label{eq20}
  G^{(0)}_{--}(p) = \frac i{p^2-m^2+i\epsilon}\,.
\end{equation}
The propagator for the $\phi_+$ field is given by
\begin{equation}\label{eq21}
  G^{(0)}_{++}(p) = \frac{-i}{p^2-m^2-i\epsilon}\,,
\end{equation}
with the $i\epsilon$ choice reflecting the fact that the $\phi_+$
originally come from complex conjugate amplitudes.  While the
Lagrangian~(\ref{eq18}) does not by itself require a $G^{(0)}_{-+}$ or
a $G^{(0)}_{+-}$ propagator, the functional integral does naturally
give such propagators because of our identification $\phi_+=\phi_-$ at
the largest time to which we evolve the system.  Such propagators also
naturally occur as Feynman lines which pass from an amplitude to a
complex conjugate amplitude in the two-time formulation of a
calculation of an expectation value or transition probability.
Clearly,
\begin{subequations}
\begin{eqnarray}
  G^{(0)}_{+-}(p) &=& 2\pi\theta(p_0)\delta(p^2-m^2)\,,\label{eq22}\\
  G^{(0)}_{-+}(p) &=& 2\pi\theta(-p_0)\delta(p^2-m^2)\,.\label{eq23}
\end{eqnarray}
\end{subequations}

Now it is straightforward to go to propagators in terms of $\phi$ and
$\pi$ fields.  For example,
\begin{subequations}
\begin{equation}\label{eq24}
  G^{(0)}_{11} = G^{(0)}_{\phi\phi} = \frac14\left[
  G^{(0)}_{--}+G^{(0)}_{++}+G^{(0)}_{-+}+G^{(0)}_{+-}\right]
\end{equation}
and
\begin{equation}\label{eq25}
  G^{(0)}_{12}=G^{(0)}_{\phi\pi}=\frac12\left[
  G^{(0)}_{--}-G^{(0)}_{++}-G^{(0)}_{-+}+G^{(0)}_{+-}\right]
\end{equation}
\end{subequations}
etc.  One easily finds
\begin{equation}\label{eq26}
\begin{split}
  G^{(0)}_{11}(p) &= \pi\delta(p^2-m^2)\,,\qquad G^{(0)}_{22}(p)=0\,,\\
  G^{(0)}_{12}(p) &= \frac i{p^2-m^2+i\epsilon p_0}\,,\qquad
  G^{(0)}_{21}(p) = \frac i{p^2-m^2-i\epsilon p_0}\,.
\end{split}
\end{equation}

\subsection{Equations for the full propagators}
\begin{figure}[ht]
\begin{center}
\def\epsfsize #1#2{0.9#1}
\epsffile{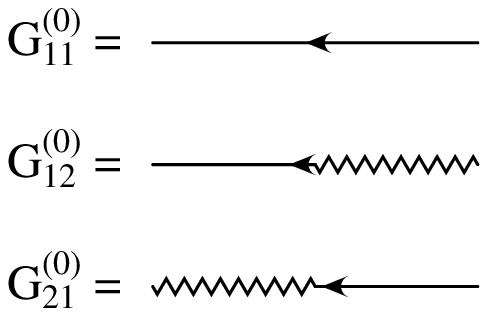}
\end{center}
\vspace{-0.25in}
\caption{The free propagators}
\label{fig1}
\end{figure}

\begin{figure}[ht]
\begin{center}
\def\epsfsize #1#2{0.9#1}
\epsffile{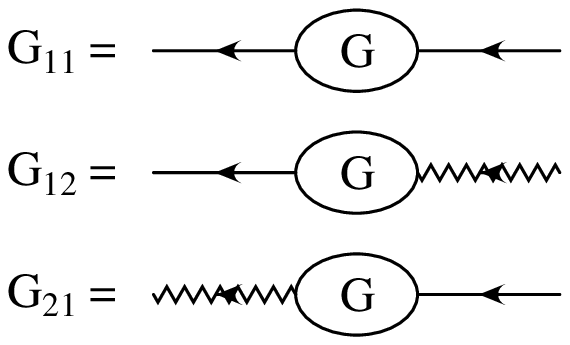}
\end{center}
\vspace{-0.25in}
\caption{The full propagators}
\label{fig2}
\end{figure}

It will be useful to have a pictorial notation for the various
propagators.  For the free propagators this notation is given in
Fig.~\ref{fig1}.  For the full propagators our pictorial notation is
given in Fig.~\ref{fig2}.  $G_{22}=0$ even in the presence of
radiative corrections because of the fact that $G_{12}$ and $G_{21}$
are causal and anticausal, respectively. (More precisely, one can
follow a $G^{(0)}_{21}$ propagator through a graph much as one can
follow an electron through a graph in QED.  In $G_{22}$ we can follow
$G^{(0)}_{21}$ propagators through the graph and they must form at
least one closed loop which is impossible because of the anticausal
property.)


\begin{figure}[ht]
\begin{center}
\def\epsfsize #1#2{0.9#1}
\epsffile{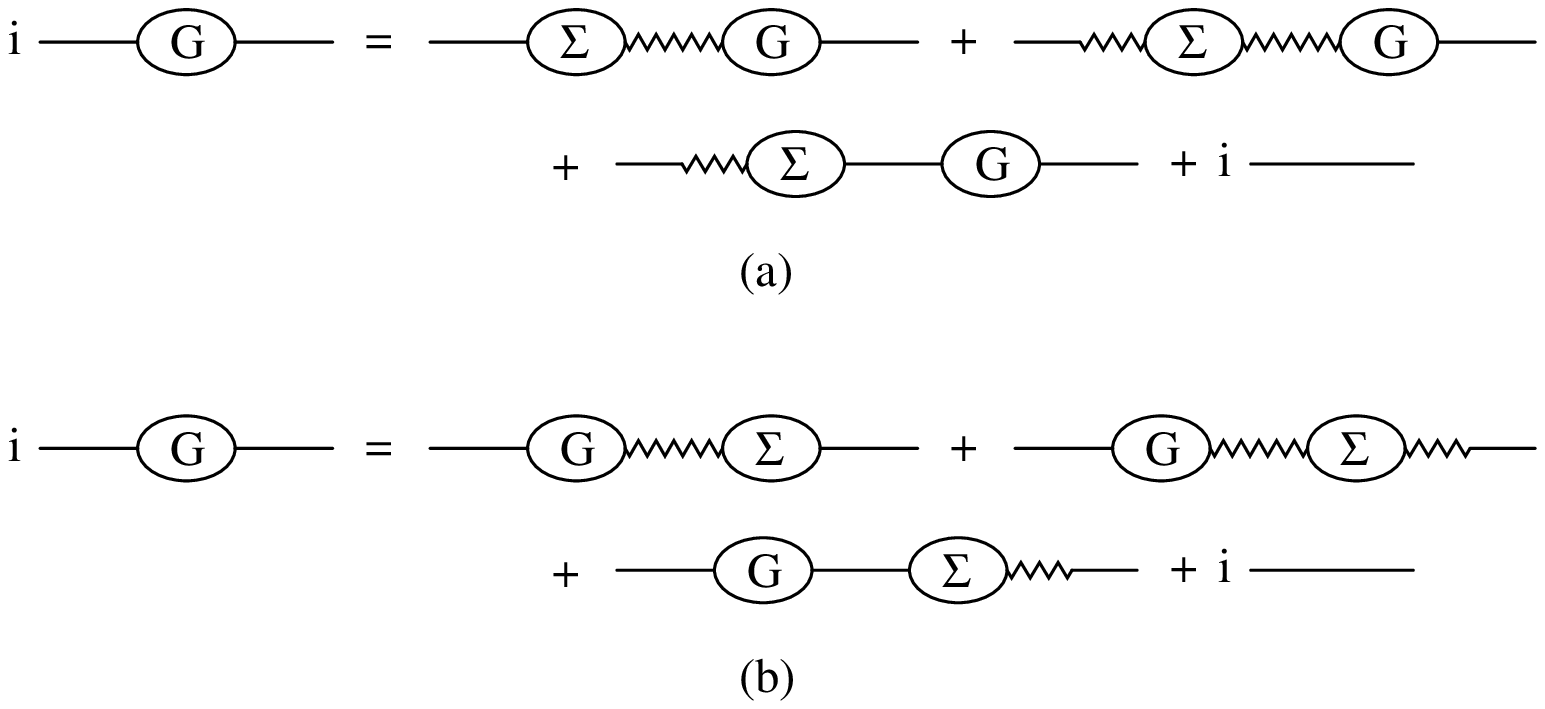}
\end{center}
\vspace{-0.25in}
\caption{Diagrammatic represetation of Eq.~(\ref{eq27-28})}
\label{fig3}
\end{figure}

One can write equations for $G$ in terms of one-particle-irreducible
parts, $\Sigma$'s, which are useful in applying perturbation theory
and in deriving a Boltzmann equation.  For $G_{11}$ this equation
takes either the form
\begin{subequations}\label{eq27-28}
\begin{equation}\label{eq27}
\begin{split}
  G_{11}(x,y) = -i\!\int\!dw\,dz\,&\left\{
  G^{(0)}_{11}(x,w)\Sigma_{12}(w,z)G_{21}(z,y) +
  G^{(0)}_{12}(x,w)\Sigma_{22}(w,z)G_{21}(z,y)\right.\\
  &\quad\left.+
  G^{(0)}_{12}(x,w)\Sigma_{21}(w,z)G_{11}(z,y)\right\} 
  + G^{(0)}_{11}(x,y)
\end{split}
\end{equation}
or
\begin{equation}\label{eq28}
\begin{split}
  G_{11}(x,y) = -i\!\int\!dw\,dz\,&\left\{
  G_{12}(x,w)\Sigma_{21}(w,z)G^{(0)}_{11}(z,y) +
  G_{12}(x,w)\Sigma_{22}(w,z)G^{(0)}_{21}(z,y)\right.\\
  &\quad\left.+
  G_{11}(x,w)\Sigma_{12}(w,z)G^{(0)}_{21}(z,y)\right\}
  + G^{(0)}_{11}(x,y)
\end{split}
\end{equation}
\end{subequations}
as illustrated in Figs.~\ref{fig3}a and Fig.~\ref{fig3}b,
respectively.

The $-i$ in Eqs.~(\ref{eq27-28}) reflects the usual
convention that $-i\Sigma$ be the sum of one-particle irreducible
graphs.  Applying $\Box_x+m^2$ to Eq.~(\ref{eq27}), $\Box_y+m^2$ to
Eq.~(\ref{eq28}) and substracting, one obtains
\begin{equation}\label{eq29}
\begin{split}
  (\Box_x-\Box_y)G_{11}(x,y) = \int\!dz\,&\left\{
   G_{11}(x,z)\Sigma_{12}(z,y) - \Sigma_{21}(x,z)G_{11}(z,y)\right.\\
  &\left.+G_{12}(x,z)\Sigma_{22}(z,y) - \Sigma_{22}(x,z)G_{21}(z,y)
  \right\}\, .
\end{split}
\end{equation}

Now it is convenient to write
\begin{equation}\label{eq30}
  G(x,y)=\int\!\frac{d^4p}{(2\pi)^4}\, G\left(\frac{x+y}2,p\right)
  e^{-ip\cdot(x-y)}
\end{equation}
with similar formalae for $\Sigma(x,y)$ in terms of
$\Sigma(\frac{x+y}2,p)$.  We now assume that the dependence of $G$ and
$\Sigma$ on $\frac{x+y}2$ is slow enough that one can replace
$\frac{x+y}2$ by $x$ in Eq.~(\ref{eq30}).  Then using
\begin{equation}\label{eq31}
  \Box_x - \Box_y = \frac{\d}{\d\left(\frac{x-y}2\right)}\cdot
  \frac{\d}{\d\left(\frac{x+y}2\right)}\,,
\end{equation}
one gets from Eq.~(\ref{eq29})
\begin{equation}\label{eq32}
  2ip\cdot\frac{\d}{\d x}G_{11}(x,p) = 
  G_{11}(x,p)[\Sigma_{21}(x,p)-\Sigma_{12}(x,p)] +
  \Sigma_{22}(x,p)[G_{21}(x,p)-G_{12}(x,p)]\, .
\end{equation}

It is important to emphasize that Eq.~(\ref{eq32}) is an equation for
a classical field theory.  We are dealing with the
Lagrangian~(\ref{eq10}) which is classical.  In the quantum case, with
$L$ given by Eq.~(\ref{eq7}), $G_{22}$ and $\Sigma_{11}$ are still
zero, but when using the equations derived in this section
one must verify that $G_{11}\gg G_{12},\,G_{21}$ so that the
approximation $\phi\gg\pi$ which leads to the classical theory is
valid.  This will, of course, depend on the initial ensemble, the
$\rho$ in Eq.~(\ref{eq16}), as well as on how far from the initial
ensemble the system has evolved.  In particular we suppose the initial
ensemble, $\rho$, is dominated by large $\phi_0$ contributions.  Of
course a system near kinetic equilibrium cannot be described by a
classical field theory.

\begin{figure}[ht]
\begin{center}
\def\epsfsize #1#2{0.9#1}
\epsffile{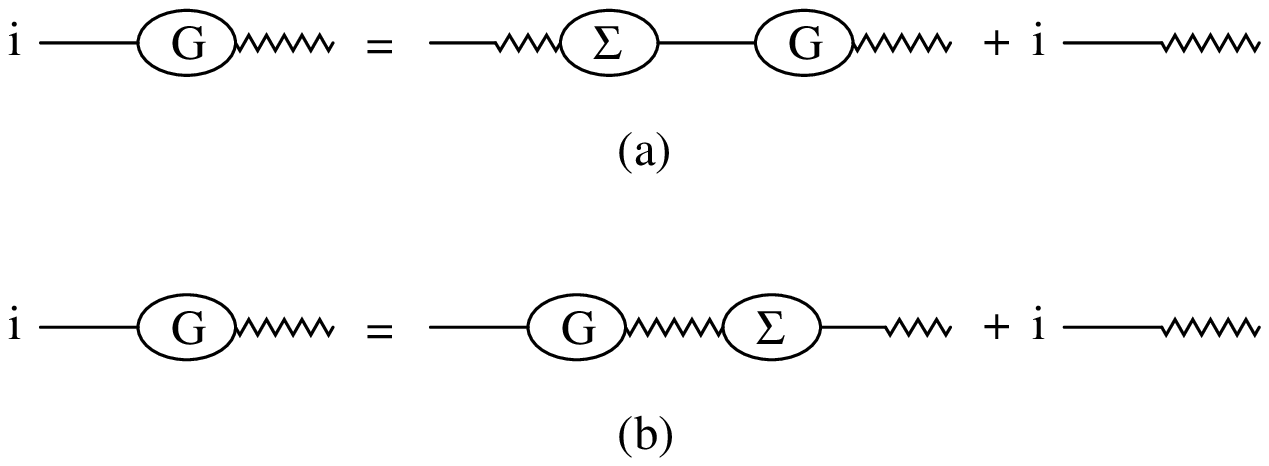}
\end{center}
\vspace{-0.25in}
\caption{Diagrammatic represetation of Eq.~(\ref{eq33-34})}
\label{fig4}
\end{figure}

Turn now to $G_{12}$.  Analogously to Eqs.~(\ref{eq27-28}) one may write
\begin{subequations}\label{eq33-34}
\begin{equation}
  G_{12}(x,y)=-i\int\!dw\,dz\,G^{(0)}_{12}(x,w)\Sigma_{21}(w.z)G_{12}(z,x)
  + G^{(0)}_{12}(x,y)
\end{equation}
and
\begin{equation}
  G_{12}(x,y)=-i\int\!dw\,dz\,G_{12}(x,w)\Sigma_{21}(w.z)G^{(0)}_{12}(z,x)
  + G^{(0)}_{12}(x,y)
\end{equation}
\end{subequations}
illustrated in Fig.~\ref{fig4}.  It is now straightforward to get
\begin{equation}\label{eq35}
  2ip\cdot\frac{\d}{\d x}G_{12}(x,p)=0
\end{equation}
with an identical equation for $G_{21}$.  Equation~(\ref{eq35}) does
not mean that the graphs shown in Fig.~\ref{fig4} do not change
$G_{12}$ from the value given in Eq.~(\ref{eq26}) for $G^{(0)}_{12}$,
but it does mean that the evolution of the system toward equilibrium
is controlled by the equation for $G_{11}$ where the right hand side
of Eq.~(\ref{eq32}) will shortly be identified with the Boltzmann
collision term.

\section{Equivalence between the Boltzmann equation and classical
field equations}
\label{sec4}

We now turn to the task of showing that the right hand side of
Eq.~(\ref{eq32}) is exactly the Boltzmann collision term.  We do this
explicitly only at lowest order in perturbation theory for the
collision term, after which the higher order corrections should be
straightforward.  We begin by identifying a part of $G_{11}$ with the
Boltzmann phase space particle density.

\subsection{The phase space density}
The generalization of Eqs.~(\ref{eq21})--(\ref{eq23}) to a medium is
straightforward.  The results are, in form, exactly like those
familiar finite temperature field thoery.  One has
\begin{subequations}\label{eq36-40}
\begin{eqnarray}
  G_{--}(x,p) &=& \frac i{p^2-m^2+i\epsilon} + 2\pi\delta(p^2-m^2)f
  \,,\stru\\
  G_{++}(x,p) &=& \frac {-i}{p^2-m^2-i\epsilon} + 2\pi\delta(p^2-m^2)f
  \,,\struu\\
  G_{-+}(x,p) &=& 2\pi\delta (p^2-m^2)[\theta(-p_0)+f]
  \,,\struu\\
  G_{--}(x,p) &=& 2\pi\delta(p^2-m^2)[\theta(p_0)+f]
  \,,
\end{eqnarray}
but in contrast to finite temperature field theory $f$ is not the
thermal phase space distribution but rather
\begin{equation}
  f = f(\x,\p, t)
\end{equation}
\end{subequations}
is a time-dependent phase space density of particles characterizing
the medium.  Equations~(\ref{eq36-40}) are
appropriate to free particles with a phase space density given by
$f$.  They are not compatible \emph{in detail} with
Eqs.~(\ref{eq27-28}) and (\ref{eq33-34}).  For
example, space and momentum dependent mass effects are not included.
Nevertheless, when $f\ll1/\lambda$ the one-loop correction to the mass
square is small compared to either the bare mass square or the square
of the typical particle momentum.  Hence
we believe Eqs.~(\ref{eq36-40}) are adequate
for our purposes of demonstrating the equivalence between classical field
theory and the Boltzmann equation
and match well with the level of accuracy of our
Boltzmann equations, Eqs.~(\ref{eq46}) and (\ref{eq47}).\footnote{For
quantities sensitive to soft modes, like the bulk viscosity, 
medium corrections to the mass are important~\cite{JeonYaffe}.}

Using equations identical to Eqs.~(\ref{eq24}), (\ref{eq25}) etc., one
easily finds
\begin{subequations}
\begin{eqnarray}
  G_{11}(x,p) &=& 2\pi\delta(p^2-m^2)\left[f+\frac12\right]
  \,,\label{eq41}\stru\\
  G_{12}(x,p) &=& \frac i{p^2-m^2+i\epsilon p_0}
  \,,\label{eq42}\stru\\
  G_{21}(x,p) &=& \frac i{p^2-m^2-i\epsilon p_0}
  \,,\label{eq43}\struu\\
  G_{22}(x,p) &=& 0\,,\label{eq44}
\end{eqnarray}
\end{subequations}
and we note that only $G_{11}$ has changed from Eq.~(\ref{eq26}).
When $f$ is large only $G_{11}$ is large, which means that $\phi$ is
large while $\pi$ is small, as we have assumed in the discussion
before Eq.~(\ref{eq10}).  Now that we have found the relationship
between the phase space density of particles, $f$, and $G_{11}$, we
can view Eq.~(\ref{eq32}) as an equation for $f$.  Indeed we can
rewrite Eq.~(\ref{eq32}) as
\begin{equation}\label{eq45}
  \left(\frac{\d}{\d t}+{\vec v}\cdot{\vec\nabla}\right)f(\x,\p,t)=
  \frac{-i}{2\omega(p)}(\Sigma_{21}-\Sigma_{12})\left(f+\frac12\right)
  +\frac i{2\omega(p)}\Sigma_{22}\,,
\end{equation}
where $p_0=\omega(p)=\sqrt{\p^2+m^2}$ is to be taken in the $\Sigma$'s
in Eq.~(\ref{eq45}).  Equation~(\ref{eq45}) has the form of a
Boltzmann equation.  Our task now is to show that the right hand side
of Eq.~(\ref{eq45}) agrees with the collision term in $\phi^4$ theory,
at least when $f\gg1$.

\subsection{The collision term in $\phi^4$ theory}
\begin{figure}[ht]
\begin{center}
\def\epsfsize #1#2{0.9#1}
\epsffile{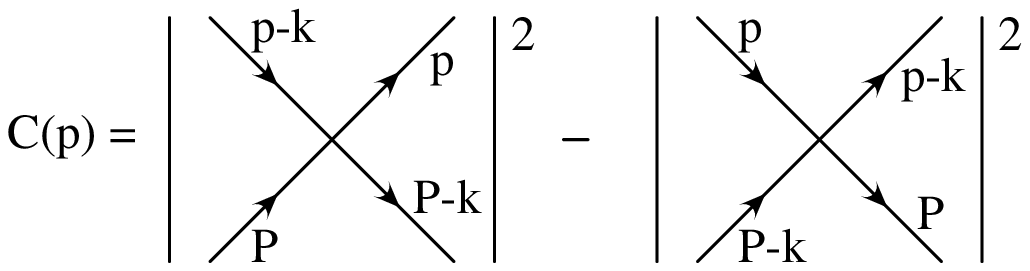}
\end{center}
\vspace{-0.25in}
\caption{The lowest order collision term}
\label{fig5}
\end{figure}

We are now going to give the lowest order contrubution to the
collision term calculated directly from the elastic scattering cross
section from the graph shown in Fig.~\ref{fig5}.  The first term on
the right hand side of Fig.~\ref{fig5} is the gain term for a particle
of momentum $\p$ while the second term is the loss term.  With the
notation
\begin{equation}\label{eq46}
  \left(\frac{\d}{\d t}+{\vec v}\cdot{\vec\nabla}\right) 
  f(\x,\p,t) = C(\x,\p,t)
\end{equation}
one has
\begin{equation}\label{eq47}
\begin{split}
  C &= \frac12\lambda^2\!\int\!
  \delta[\omega(p{-}k){+}\omega(P){-}\omega(p){-}\omega(P{-}k)]
  \frac{d^3P\,d^3k}{(2\pi)^52\omega(p{-}k)2\omega(P)2\omega(p)2\omega(P{-}k)}
  \times\stru\\
  &\quad \times\Bigl\{f(p{-}k)f(P)[1+f(p)][1+f(P{-}k)] -
  f(p)f(P{-}k)[1+f(p{-}k)][1{+}f(P)]\Bigr\}
\end{split}
\end{equation}

The first factor on the right hand side of Eq.~(\ref{eq47}), the
$\frac12$, is a symmetry factor.  The term quartic in the $f$'s
cancels in Eq.~(\ref{eq47}) and one is left with cubic and quadratic
terms.  The cubic term is
\begin{equation}\label{eq48}
  \{~\}_3=  f(p{-}k)f(P)[f(p)+f(P{-}k)]-f(p)f(P{-}k)[f(p{-}k)+f(P)]
\end{equation}
with the first two terms on the right hand side of Eq.~(\ref{eq48})
coming from the gain term and the second two terms coming from the
loss term.  The quadratic term is
\begin{equation}\label{eq49}
  \{~\}_2 =  f(p{-}k)f(P)-f(p)f(P{-}k)
\end{equation}
We shall see that the classical field evolution reproduces the
$\{~\}_3$ part of the collision term and, perhaps surprisingly, even
the $\{~\}_2$ part.

\subsection{Evaluating the collision term from classical field theory}
\begin{figure}[ht]
\begin{center}
\def\epsfsize #1#2{0.9#1}
\epsffile{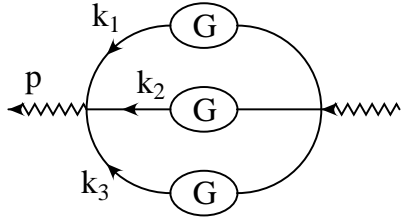}
\end{center}
\vspace{-0.25in}
\caption{Lowest order contribution to $\Sigma_22$}
\label{fig6}
\end{figure}

Now we turn to evaluating the right hand side of Eq.~(\ref{eq45}) at
lowest order, order $\lambda^2$, in perturbation theory.  We begin
with the $\Sigma_{22}$-term whose lowest order contribution is
illustrated in Fig.~\ref{fig6}.  One has
\begin{equation}\label{eq50}
  \frac{i\Sigma_{22}}{2\omega(p)}=\frac i{2\omega(p)} i
  \frac{(-i\lambda)^2}{3!}\int\!\frac{d^4k_1\,d^4k_2\,d^4k_3}{(2\pi)^8}\,
  \delta^4(k_1+k_2+k_3-p)\,G_{11}(k_1)G_{11}(k_2)G_{11}(k_3)\,,
\end{equation}
where the second factor of $i$ on the right hand side of
Eq.~(\ref{eq50}) comes because $-i\Sigma_{22}$ is given by the usual
Feynman rules, and the $1/3!$ is the symmetry factor for the graph.
In one of the three factors of $G_{11}$ we take $k_0<0$ while in the
remaining two factors $k_0>0$.  (This is the only way to satisfy the
$\delta$-function constraint.  The choice of one of the lines to have
$k_0<0$ also introduces a counting factor of 3.)  Using
Eq.~(\ref{eq41}), it is now straightforward to get
\begin{equation}\label{eq51}
\begin{split}
  \frac{i\Sigma_{22}}{2\omega(p)} = \frac{\lambda^2}2&\!\int\!
  \delta[\omega(p{-}k){+}\omega(P){-}\omega(p){-}\omega(P{-}k)]
  \frac{d^3P\,d^3k}{(2\pi)^52\omega(p{-}k)2\omega(P)2\omega(P{-}k)2\omega(p)}
  \\
  &\quad \left[f(p{-}k)+\half\right]\left[f(P)+\half\right]
  \left[f(P{-}k)+\half\right]\,,
\end{split}
\end{equation}
where we have chosen, say, $k_1=P$, $k_2=p-k$, $k_3=-(P-k)$, to match
the picture of incoming and outgoing lines in Fig.~\ref{fig5}.  One
easily sees that the cubic term in $f$ agrees with one of the gain
terms in Eq.~(\ref{eq47}).

\begin{figure}[ht]
\begin{center}
\def\epsfsize #1#2{0.9#1}
\epsffile{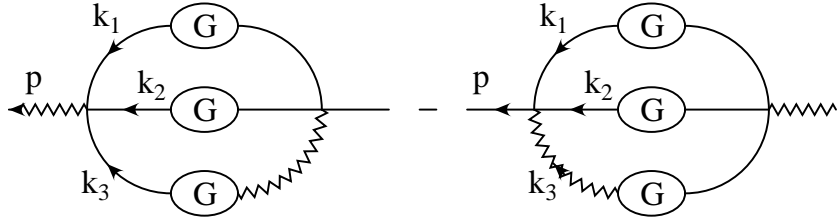}
\end{center}
\vspace{-0.25in}
\caption{$\Sigma_{21}-\Sigma_{12}$ at lowest order}
\label{fig7}
\end{figure}

Now we turn to the remaining term on the right hand side of
Eq.~(\ref{eq45}), with $\Sigma_{21}-\Sigma_{12}$ illustrated in
Fig.~\ref{fig7}, at order $\lambda^2$.  One has
\begin{equation}\label{eq52}
\begin{split}
  \frac{-i}{2\omega(p)}\left[f(p)+\half\right](\Sigma_{21}-\Sigma_{12})
 &=\frac{-i}{2\omega(p)}i\frac{(-i\lambda)^2}2\!\int\!
  \frac{d^4k_1\,d^4k_2\,d^4k_3}{(2\pi)^8}\,\delta^4(k_1+k_2+k_3-p)
  \\
 &\quad 2\pi\epsilon(k_{30})\delta(k_3^2-m^2)G_{11}(k_1)G_{11}(k_2)
\end{split}
\end{equation}

When $k_{30}>0$ there are two (identical) terms, one having $k_{10}>0$
and $k_{20}<0$ and the other having $k_{10}<0$ and $k_{20}>0$.  When
$k_{30}<0$ it is necessary that $k_{10}$ and $k_{20}$ both be
positive.  Thus one finds
\begin{equation}\label{eq53}
\begin{split}
  \frac{-i}{2\omega(p)}\left[f(p)+\half\right](\Sigma_{21}&-\Sigma_{12})
  =-\frac{\lambda^2}2\!\int\!
  \delta[\omega(P{-}k)+\omega(P)-\omega(p)-\omega(P{-}k)]\\
  &\frac{d^3P\,d^3k}{(2\pi)^52\omega(p)2\omega(p-k)2\omega(P)2\omega(P-k)}
  \left[f(p)+\half\right]\times\\
  &\times\left\{2\left[f(P{-}k)+\half\right]\left[f(P)+\half\right]-
         \left[f(p{-}k)+\half\right]\left[f(P)+\half\right]\right\}
\end{split}
\end{equation}
where in the first term in $\{\,\}$ in Eq.~(\ref{eq53}) we have taken
$k_3=p-k$, $k_1=-(P-k)$, $k_2=P$ along with $k_1\leftrightarrow k_2$,
while in the second term we have taken $k_1=(p{-}k)$, $k_2=P$, and
$k_3=-(P-k)$ so that the variables match those in Eq.~(\ref{eq48}).
We note that the two loss terms in Eq.~(\ref{eq48}) are in fact
identical.

Now, taking the terms cubic in $f$ in Eqs.~(\ref{eq51}) and
(\ref{eq52}) exactly reproduces the cubic term, (\ref{eq48}), in
Eq.~(\ref{eq47}).  For the quadratic terms we find, from
Eqs.~(\ref{eq51}) and (\ref{eq53}), the result that replaces $\{~\}_2$
in Eq.~(\ref{eq49}) is
\begin{equation}\label{eq54}
\begin{split}
  \{~\}'_2 = \frac12 \Bigl\{& f(p{-}k)f(P) + f(p{-}k)f(P{-}k) + f(P)f(P{-}k)
  - 2f(p)f(P{-}k)\\ 
  &- 2f(p)f(P)
  -2f(P{-k})f(P) + f(p)f(p{-}k) + f(p)f(P) + f(p{-}k)f(P)\Bigr\}
\end{split}
\end{equation}
Using the fact that $f(p{-}k)f(P{-}k)$ and $f(P)f(P{-}k)$ are
equivalent expressions, after integration in Eq.~(\ref{eq51}) or
(\ref{eq53}), as are $f(p)f(P)$ and $f(p)f(p-k)$ one finds
$\{~\}'_2=\{~\}_2$.  Finally there are the linear terms in the $f$'s
coming from Eqs.~(\ref{eq51}) and (\ref{eq53}) for which there are no
counterparts in Eq.~(\ref{eq47}).  In fact the linear terms do not
cancel in Eq.~(\ref{eq51}) and (\ref{eq53}) so that the classical
field theory does not \emph{exactly} reproduce the Boltzmann collision
term.

Finally, a comment on the level of accuracy at which one can expect
the classical field theory to reproduce the Boltzmann equation.  We
always suppose that $f\gg1$ but that $\lambda f\ll1$ so that a
perturbative discussion makes sense.  The size of our fields then are
$\phi\sim\sqrt f$ while $\pi\sim1/\sqrt f$.  In going from the full
Lagrangian, given in Eq.~(\ref{eq7}), to the classical Lagrangian,
given in Eq.~(\ref{eq10}), we have dropped a term
$-\frac\lambda{3!}\pi^3\phi$ compared to the interaction term
$-\frac\lambda{3!}\pi\phi^3$ which has been kept.  Now an estimate of
the size of these two terms is
\begin{equation}\label{eq55}
  \lambda\pi\phi^3\sim\lambda f
\end{equation}
and
\begin{equation}\label{eq56}
  \lambda\pi^3\phi\sim\frac\lambda f
\end{equation}
Thus, the quantum corrections are naturally down by a factor of
$1/f^2$ as compared to a classical evaluation at a corresponding order
of $\lambda$.  Since the maximal possible terms in the collision term
are of size $\lambda^2f^3$, one expects quantum corrections to occur
at a level of $\lambda^2f$ leaving the $\lambda^2f^2$ terms still in
the classical field theory domain.  So in retrospect, our ability to
get both the $\lambda^2f^3$ and $\lambda^2f^2$ terms in
Eq.~(\ref{eq47}) correctly was to be expected.  This means that
the $2\to2$ parts of a collision term can be obtained from a classical
theory calculation.  This cannot be expected to be the case for
$2\to4$ processes where terms of size $\lambda^4f^5$, $\lambda^4f^4$,
$\lambda^4f^3$, and $\lambda^4f^2$ in the collision terms will occur
at leading order in $\lambda$.  One can here expect to get the
$\lambda^4f^5$ and $\lambda^4f^4$ terms correctly in a classical field
theory, but not the $\lambda^4f^3$ and $\lambda^4f^2$ terms.

\acknowledgments 

We thank Dietrich B\"odeker for encouraging us to write up this paper.
The work of A.H.M. is supported, in part, by a DOE grant.  D.T.S. is
supported, in part, by DOE grant No.\ DOE-ER-41132 and by the Alfred
P.\ Sloan Foundation.

\end{document}